\documentclass{WileyMSP-template}
\usepackage{lineno}
\usepackage{todonotes}
\usepackage{amsmath} 
\usepackage{amsfonts}
\usepackage[capitalise]{cleveref}
\justifying

\begin{document}


\title{Vectorial engineering of second-harmonic generation in silicon-based waveguides integrated with 2D materials}

\maketitle


\author{Mohd Rehan$^{\S}$}
\author{Nathalia B. Tomazio$^{\S}$}
\author{Alisson R. Cadore}
\author{Daniel F. Londono-Giraldo}
\author{Daniel A. Matos}
\author{Gustavo S. Wiederhecker}
\author{Christiano J. S. de Matos$^*$}


\begin{affiliations}
$\S$ Authors who contributed equally to this work.

M. Rehan, D. F. Londono-Giraldo, C. J. S. de Matos\\
School of Engineering, Mackenzie Presbyterian University, Sao Paulo 01302-907, SP, Brazil\\
MackGraphe, Mackenzie Presbyterian Institute, Sao Paulo 01302-907, SP, Brazil\\
Email Address: cjsdematos@mackenzie.br

N. B. Tomazio, D. A. Matos\\
Instituto de Física, Universidade de São Paulo, São Paulo, Brazil

A. R. Cadore\\
Brazilian Nanotechnology National Laboratory (LNNano), Brazilian Center for Research in Energy and Materials (CNPEM), Campinas 13083-100, SP, Brazil\\
Programa de Pós-Graduação em Física, Instituto de Física, Universidade Federal de Mato Grosso,
Cuiabá 78060-900, MT, Brazil

G. S. Wiederhecker\\
Photonics Research Center, Gleb Wataghin Physics Institute, Universidade Estadual de Campinas (UNICAMP), Campinas, SP, Brazil
\end{affiliations}



\begin{abstract}
Integrating 2D materials onto on-chip photonic devices holds significant potential for nonlinear frequency conversion across various applications. The lack of inversion symmetry in monolayers of transition metal dichalcogenides (TMDs), e.g.,  MoS$_2$, is particularly attractive for enabling nonlinear phenomena based on $\chi^{(2)}$ in silicon-based photonic devices incorporated with these materials, which has been previously demonstrated. However, reports have largely overlooked the need to consider, in the nonlinear modal interaction, both the tensorial nature of the TMD's second-order susceptibility and the full vectorial nature of the electromagnetic fields. In this work, we investigate second-harmonic generation (SHG) in silicon nitride (SiN) waveguides integrated with a monolayer of MoS$_2$.
We experimentally observed an enhancement in SHG in MoS$_2$-loaded waveguides compared to those without the monolayer. Notably, this enhancement occurred even when the dominant electric field component of the pump and/or signal mode was orthogonal to the TMD plane, highlighting co- and cross-polarized SHG interactions. This phenomenon cannot be predicted by the traditionally used scalar models.
By taking into account the full vectorial and tensorial natures of the problem, we then designed a waveguide in which a TE pump mode is phase-matched to a TM second-harmonic mode. With a single 110-µm-long MoS$_2$ flake, we experimentally achieved $14\times$ frequency conversion enhancement relative to the non-phase-matched case and $220\times$ enhancement relative to free-space (normal-incidence) excitation. Our work, thus, introduces fundamental guidelines for the design and optimization of nonlinear silicon-photonic devices based on 2D-material hybrid integration. These guidelines are material independent and may lead to significant further conversion efficiency enhancement. 
\end{abstract}

\section{Introduction}
Integrated photonic devices are poised to leverage nonlinear optical phenomena, particularly those derived from second-order nonlinear susceptibility ($\chi^{(2)}$). These phenomena include second harmonic generation (SHG), parametric down-conversion, and difference frequency generation. Such devices provide a versatile platform for various applications~\cite{thomson2016roadmap}, including frequency conversion~\cite{wang2018ultrahigh, hwang2023tunable}, electro-optic modulation~\cite{xiong2012aluminum}, and quantum information processing~\cite{harper2024highly, guo2017parametric, gray2024large}. 

Silicon-based materials, although attractive due to their mature fabrication technology and CMOS compatibility, inherently lack bulk $\chi^{(2)}$ nonlinearity due to their centrosymmetric crystal structure. Several strategies have been developed to enable nonlinear phenomena based on $\chi^{(2)}$ in silicon photonics, such as strain~\cite{jacobsen2006strained} and electric field-induced symmetry breaking ~\cite{timurdogan2017electric} and hybrid integration~\cite{rao2018second, pelgrin2023hybrid}. 
These strategies add manufacturing complexity to the device, limiting its scalability. In fact, applying electric fields requires embedding electrodes in the device, while adding a stressor layer to the waveguide for strain-induced symmetry breaking can constrain phase matching engineering~\cite{timurdogan2017electric}. Hybrid integration with Lithium Niobate (LN) or other strong $\chi^{(2)}$-based materials shows promising results~\cite{rao2018second}, but also suffers from design and manufacturing complexity since it involves mode coupling from the silicon-based to the LN-based waveguide, in addition to periodic poling of the LN to ensure quasi-phase matching~\cite{jankowski2020ultrabroadband}. 
More recently, the integration of chip-scale silicon photonic devices with layered group-VI transition metal dichalcogenides (TMDs), such as molybdenum disulfide (MoS$_2$), tungsten disulfide (WS$_2$) and others, has been investigated for this purpose~\cite{liu2022silicon, kuppadakkath2022direct, HeNL2021, Javerzac-GalyNL2018}. 

TMDs in the 2H phase with a single layer or an odd number of layers exhibit second-order susceptibility, arising from the absence of inversion symmetry in their crystalline structure~\cite{autere2018nonlinear, MalardPRB213, HamzaJAP2022, ViannaNano2021}. These materials have been shown to exhibit large effective bulk second-order susceptibilities [$\chi^{(2)}$ $\sim$ 10$^{-10}$ - 10$^{-11}$ mV$^{-1}$]~\cite{pelgrin2023hybrid, woodward2016characterization}, offering new avenues for nonlinear optical applications. In addition to their compact size, they can be directly transferred~\cite{HeNL2021, Javerzac-GalyNL2018, liu2015enhanced, chen2017enhanced} or grown~\cite{liu2022silicon} on top of integrated silicon devices without the need for lattice matching~\cite{xia2014two}, which represents a convenient alternative to the challenges of heterogeneous integration with LN or III-V semiconductors~\cite{rao2018second, rabiei2013heterogeneous,xiong2011integrated}. 
Moreover, by interfacing waveguide and cavity-based devices with TMDs, previous works have demonstrated that the interaction length of light with the TMD material can be significantly extended, allowing one to overcome the low single-pass conversion efficiency attributed to its atomic thickness~\cite{fryett2016silicon,chen2017enhanced, rarick2024enhanced, fujii2024van}. 
Several device architectures have demonstrated substantial SHG enhancement, including Fabry–Pérot microcavities incorporating MoS$_2$ (10$\times$ enhancement) \cite{day2016microcavity}, silicon waveguides integrated with monolayer MoSe$_2$ (280$\times$ enhancement) \cite{chen2017enhanced}, and silicon photonic-crystal cavities with WSe$_2$ (200$\times$ enhancement) \cite{fryett2016silicon}. 
These enhancements were achieved when compared to the free-space excitation of the monolayer TMD at normal incidence on a corresponding homogeneous substrate. Substrate-assisted configurations have also been explored, such as MoS$_2$ and WS$_2$ monolayers on epsilon-near-zero (ENZ) substrates, yielding a 10$\times$ enhancement \cite{ViannaNano2021}. Collectively, these studies highlight the versatility of TMDs for nonlinear optical conversion in integrated platforms and motivate the investigation of waveguiding geometries and modeling approaches to further improve efficiency and physical understanding.

\begin{figure}
  \includegraphics[width=\linewidth]{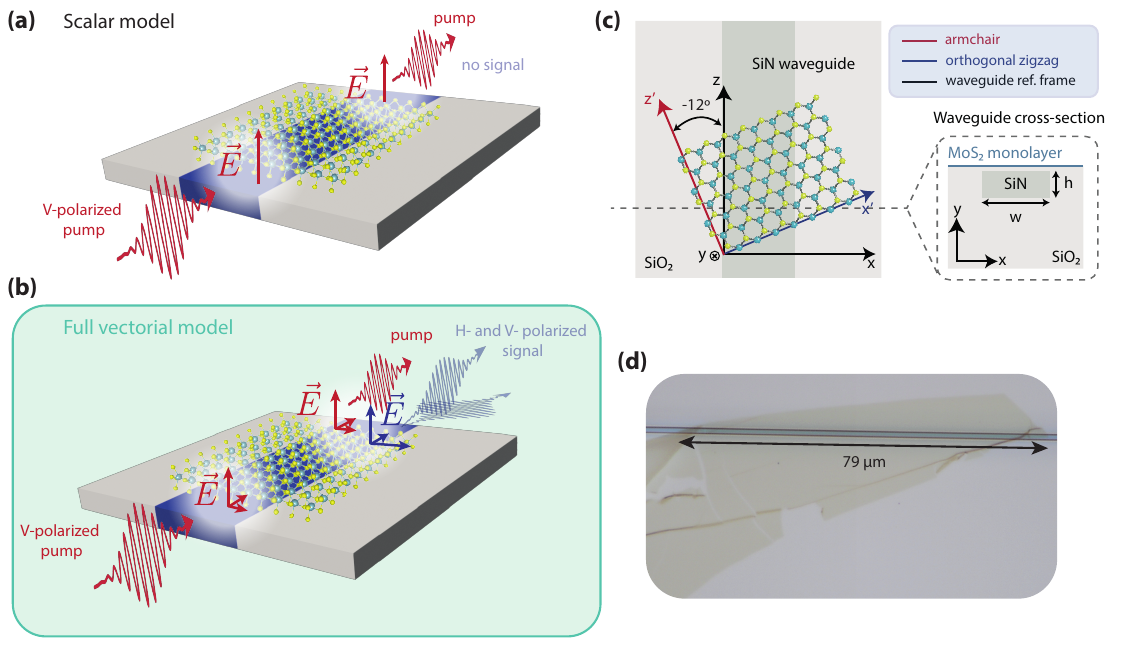}
  \caption{\textbf{MoS$_2$-loaded waveguide: models, design and 2D material transfer:} (a-b) Schematics of SHG in the MoS$_2$-loaded waveguides as described by the (a) the scalar model and (b) the full vectorial model. Since the scalar model only considers the main electric field component of the modes and the 2D material does not interact with out-of-plane electric fields, according to this model, there is no SHG for vertical-polarization pumping. In contrast, the vectorial model predicts SHG at horizontal and vertical polarizations for the vertical-polarized pump. 
  (c) Schematics of the SiN WG integrated with a MoS$_2$ monolayer. The WG is cladded in a SiO$_2$ substrate and has cross-section dimensions of 1 $\mu$m $\times$ 0.8 $\mu$m (width $\times$ height). There is a 100 nm spacing between the WG top surface and the MoS$_2$ flake. Also shown is the MoS$_2$ ($x'$, $z'$) and the WG ($x$, $z$) reference frames, highlighting the angle $\theta$ between them. (d) Optical micrograph of the 79 $\mu$m-long MoS$_2$ transferred onto the SiN WG.} 
  \label{fig:design_considerations}
\end{figure}

Although previous work have demonstrated second-order nonlinearities in on-chip Si-based devices through integration with TMDs~\cite{chen2017enhanced, liu2022silicon}, the complete nonlinear modal interaction, jointly accounting for the full vectorial nature of the involved fields and the tensorial nature of the second-order susceptibility of the TMD, has not been fully described. The adopted scalar models neglect the electric field components along the waveguide axis, which play a fundamental role in cross-polarized and copolarized modal interactions. With such a simplistic, yet widely used, approach, nonlinear interaction would be negligible with modes for which the main electric field component is orthogonal to the 2D material (see \cref{fig:design_considerations}(a-b)). The full vectorial-tensorial model is required to properly determine the second harmonic (SH) signal modes for which the nonlinear interaction occurs more efficiently, providing crucial guidelines for design engineering towards nonlinear frequency conversion.   %

In this work, we demonstrate SHG at 780 nm in silicon nitride (SiN) waveguides integrated with a dry-transferred MoS$_2$ monolayer, as illustrated in \cref{fig:design_considerations}(c-d). We provide an experimental characterization of SHG, which demonstrates an increase in SH power in waveguides (WGs) incorporating a monolayer of MoS$_2$, compared to those without TMD integration.
We provide a detailed analysis of SH conversion, focusing on the vectorial/tensorial nature of the nonlinear interaction and on how the conversion efficiency depends on the angle between the armchair direction of the MoS$_2$ flake and the WG axis. Our results demonstrate that, despite the absence of out-of-plane $\chi^{(2)}$ components in the TMD, the highest conversion efficiency can be achieved with the pump and/or signal modes whose main electric field component is orthogonal to the 2D material (i.e., transverse magnetic, TM, modes). 
We focus our study on monolayer TMDs because they constitute the dominant and most systematically reported class of 2D materials for on-chip SHG enhancement in waveguide and cavity platforms; comparable on-chip datasets for other 2D materials remain limited. Nevertheless, our approach is general and must also be applied to other 2D materials, if accurate frequency conversion modeling/optimization is to be achieved. By applying the model and tailoring a waveguide for modal phase matching, we achieve $220 \times$ SHG enhancement relative to free space pumping.

\section{Second harmonic generation in  MoS$_2$-loaded waveguides}

The MoS$_2$-loaded WG is a SiN WG integrated with a MoS$_2$ monolayer, as illustrated in \cref{fig:design_considerations}(c-d). In the bare SiN waveguide, second-order nonlinearities occur at the interface between core and cladding and cannot be controlled. In contrast, the MoS$_2$ monolayer is non-centrosymmetric, which enables second-order nonlinear phenomena in the MoS$_2$-loaded WG ~\cite{autere2018nonlinear, MalardPRB213, HamzaJAP2022, ViannaNano2021}. The modes guided in the WG interact with the MoS$_2$ crystal on the WG top surface via their evanescent field. The interaction length between the WG and the transferred MoS$_2$ flake is 79 $\pm$ 3 $\mu$m (\cref{fig:design_considerations}(d)), and the angle between the WG axis and the MoS$_2$ armchair direction is -12 degrees (Fig. S1(a)). Raman measurements revealed the E$^1_{2g}$ mode at $\sim$ 384 cm$^{-1}$ and the A$_{1g}$ mode at $\sim$ 403 cm$^{-1}$, with a characteristic spacing of 19 cm$^{-1}$ ~\cite{XuMoS2Raman}, confirming the monolayer nature of the MoS$_2$ flake (Fig. S1(c)). See \textit{Methods} for details.

\begin{figure}
  \includegraphics[width=\linewidth]{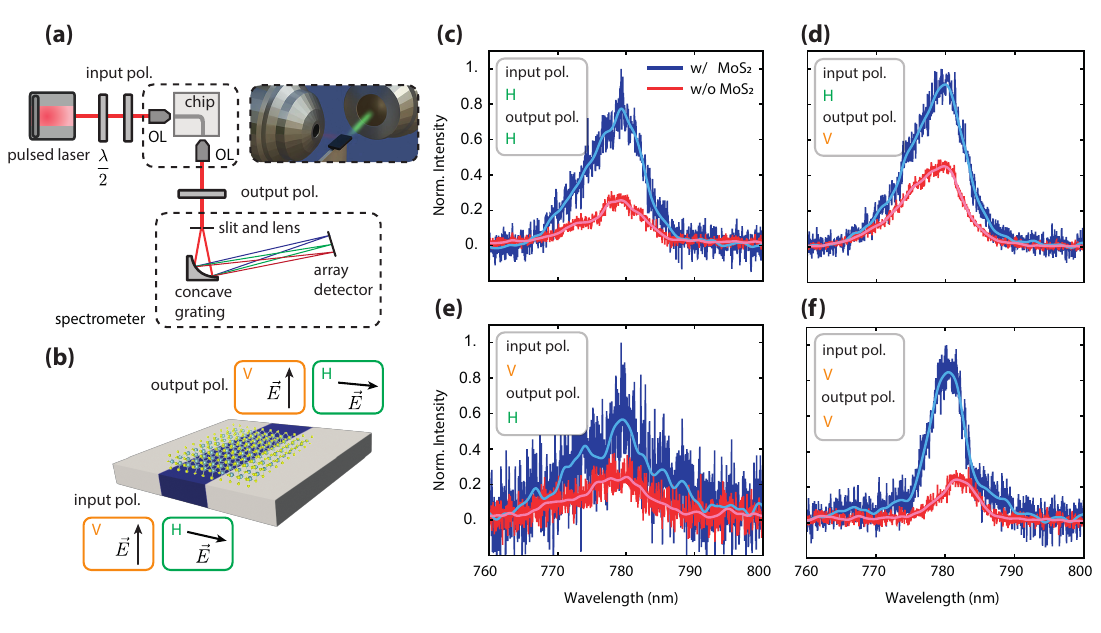}
  \caption{\textbf{Experimental results of SHG in the MoS$_2$-loaded waveguides:} (a) Main components of the experimental setup. The laser is an Erbium-doped fiber laser @ 1560 nm, with 150 fs of time duration and 89 MHz of repetition rate, delivering 750 W of input (off-chip) peak power. $\lambda/2$ stands for half-wave plate. The inset shows a 3D view of the coupling region, with the input and output objective lenses and the chip. See \textit{Methods} for further details. (b) Schematics of the input and output polarization states. H and V are defined with respect to the chip plane. (c-f) Normalized SH-intensity collected at the chip output for WGs with MoS$_2$ (blue) and without it (red) at different polarization configurations, as indicated in the subplot legends. The light blue and light red curves represent low-pass filtered curves. To compare the signals with and without MoS$_2$ in each subplot, the curves are normalized by the square of the pump power collected at the chip output, eliminating the influence of insertion losses in the chip. For visual clarity, within each subplot, the curves are further normalized to the maximum SH intensity of the MoS$_2$-loaded waveguide.}
  \label{fig:SHG_experiment}
\end{figure}

To demonstrate the feasibility of second-order nonlinear phenomena enabled by incorporating MoS$_2$ onto the SiN WG, we investigated the SHG response. The experimental characterization of SHG in waveguides with and without MoS$_2$ is shown in \cref{fig:SHG_experiment}. The WGs were pumped at 1560 nm with fs-pulses with 750 W of input (off-chip) peak power and the pump polarization was adjusted to either vertical (V) or horizontal (H) using free-space polarizers and wave plates.
(\cref{fig:SHG_experiment}(a-b)). The WG is single-mode at the pump frequency but highly multimode at the SH-wavelength (780 nm) ($\sim 16$ supported modes). The output light passes through a polarizer and is directed to a spectrometer. A second free-space polarizer at the output selects the modes contributing to SH-power along the V or H polarizations. Following our convention, the TE (TM) modes exhibit the highest electric field component along the WG width (height), and the subscript refers to the mode order (ranked from the highest to the lowest modal effective index) for each polarization. \cref{fig:SHG_experiment}(c-f) shows the SHG spectra for WGs with and without the MoS$_2$ monolayer at different polarization configurations. For input H polarization, which sets the pump field in the WG to the TE$_0$ mode, we observe an enhancement of the SHG in the WGs with MoS$_2$ flake of $2.4$ and $1.9$ for output H and V polarizations, respectively. For input V-polarization, corresponding to the TM$_0$ mode as the pump field, the SHG enhancement in the WGs with MoS$_2$ is $2.2$ and $3.1$ for the output H- and V-polarizations, respectively.
The enhancement factors were calculated by taking the ratio between the integrated spectra with and without MoS$_2$. The smaller SH power observed for the WGs without MoS$_2$ (red spectra of \cref{fig:SHG_experiment}(c-f)) arises from the centro-symmetry-breaking at the WG SiN-to-SiO$_2$ and SiO$_2$-to-air interfaces \cite{Levy:11, bloembergen1968optical, tom1983second, shen1989surface}.

To investigate the physics behind the SH-enhancement factors, especially in the case of the pump and/or signal modes with polarization orthogonal to the MoS$_2$ monolayer,   
we performed simulations based on nonlinear coupled equations \cite{boyd2020nonlinear,shen1984principles}, which are described in detail in Supplementary Information section S1. The SH conversion efficiency ($\eta$) depends on the WG length ($L$), the phase-mismatch factor ($\Delta\beta$) and the nonlinear coefficient ($\gamma$) between pump and signal modes: 

\begin{equation}
    \eta  = \gamma^2\,L^2\,\frac{\sin^2(\Delta\beta\,L/2)}{(\Delta\beta\,L/2)^2}\;.
    \label{eq:conversion_eff_main}
\end{equation}

\noindent The phase mismatch factor, $\Delta\beta= 2\beta_P-\beta_S$, with $\beta_P$ and $\beta_S$, respectively, representing the pump and signal propagation constants, may deviate from zero, causing the SH-power to oscillate along the WG. $\gamma$ measures the nonlinear interaction strength, which is proportional to the spatial overlap between the pump and signal mode components ($\vec{e_P}$ and $\vec{e_S}$) in the nonlinear medium, coupled by the second-order susceptibility tensor ($\stackrel{\leftrightarrow}{\chi}^{(2)}$):

\begin{equation}
	\gamma = \frac{\,\omega_{S}\,\varepsilon_0}{4}\,\left[\int_{MoS_2} \vec{e}\,^*_{S} \cdot \stackrel{\leftrightarrow}{\chi}^{(2)}:\,\vec{e}_P\,\vec{e}_P\,da\right]\,=
    \frac{\,\omega_{S}\,\varepsilon_0\,\chi^{(2)}}{4}\,\left[\int_{MoS_2}\Omega\,da \right]\,.
	\label{eq:gamma_main}
\end{equation}


\noindent In \cref{eq:gamma_main}, $\omega_S$ and $\varepsilon_0$ represent, respectively, the angular frequency of the signal and the vacuum permittivity. The variable $\Omega$ captures the pump and signal field overlap terms and the dependence of the SHG interaction with the angle $\theta$ between the MoS$_2$ armchair direction and the WG axis, reflecting the 3-fold rotational symmetry of the MoS$_2$ crystal \cite{chen2017enhanced}:

\begin{equation}
\begin{split}
\Omega = \,&[e_{S,z}^*~e_{P,z}^2 - e_{S,z}^*~e_{P,x}^2 - 2\,e_{S,x}^*~e_{P,z}^{}~e_{P,x}^{}]\,cos(3\theta)\,+\\
            &[e_{S,x}^*~e_{P,z}^2 - e_{S,x}^*~e_{P,x}^2 + 2\,e_{S,z}^*~e_{P,z}^{}~e_{P,x}^{}]\,sin(3\theta).
\end{split}
\label{eq:angleDependence_main}
\end{equation}

\noindent In \cref{eq:angleDependence_main}, $x$ and $z$ are the coordinates illustrated in \cref{fig:design_considerations}(c). 
Since the MoS$_2$'s $\chi^{(2)}$ does not couple the pump and signal mode components along the out-of-plane direction ($y$), one would not expect SHG involving TM pump/signal modes, whose main electric field component is along $y$. However, these modes can interact through their electric field component along the waveguide axis ($z$), which has the same weight as the $x$-field component - the primary field component of a TE mode - in the spatial overlap factor (\cref{eq:angleDependence_main}). Therefore, accounting for the vectorial nature of the electric fields is essencial, especially to describe SHG involving TM $\rightarrow$ TM and cross-polarized (TE $\rightarrow$ TM and TM $\rightarrow$ TE) modal interactions.   
Note that the overlap integral is taken over the MoS$_2$ region, which is the material contributing to SHG in the MoS$_2$-loaded WGs. Therefore, the signal modes with the highest nonlinear coefficient are the ones that overlap the most with the pump mode in the MoS$_2$ domain.  

It is important to mention that the formalism presented herein is entirely general and can be directly applied without modification to other 2H-phase transition-metal dichalcogenides (TMDs) that share the same D$_{3h}$ crystal symmetry, such as MoSe$_2$, WS$_2$ and WSe$_2$, by simply substituting their respective material parameters and $\chi^{(2)}$ magnitudes. For other noncentrosymmetric 2D materials with different symmetries (for example, ReS$_2$ or elemental tellurium), the same vectorial–tensorial model remains valid once the appropriate $\chi^{(2)}$ tensor components are used.
Furthermore, the full-vectorial framework developed here provides a natural foundation for describing higher-order nonlinearities. The approach can naturally be extended to $\chi^{(3)}$-mediated processes such as third-harmonic generation and four-wave mixing—in 2D-material-loaded waveguides, by incorporating the corresponding third-order polarization terms into Maxwell’s equations. Hence, the present formulation constitutes a unified and versatile model for predicting nonlinear interactions of arbitrary order in hybrid photonic structures integrating 2D materials.

\begin{figure}
    \centering
  \includegraphics[width=.9\linewidth]{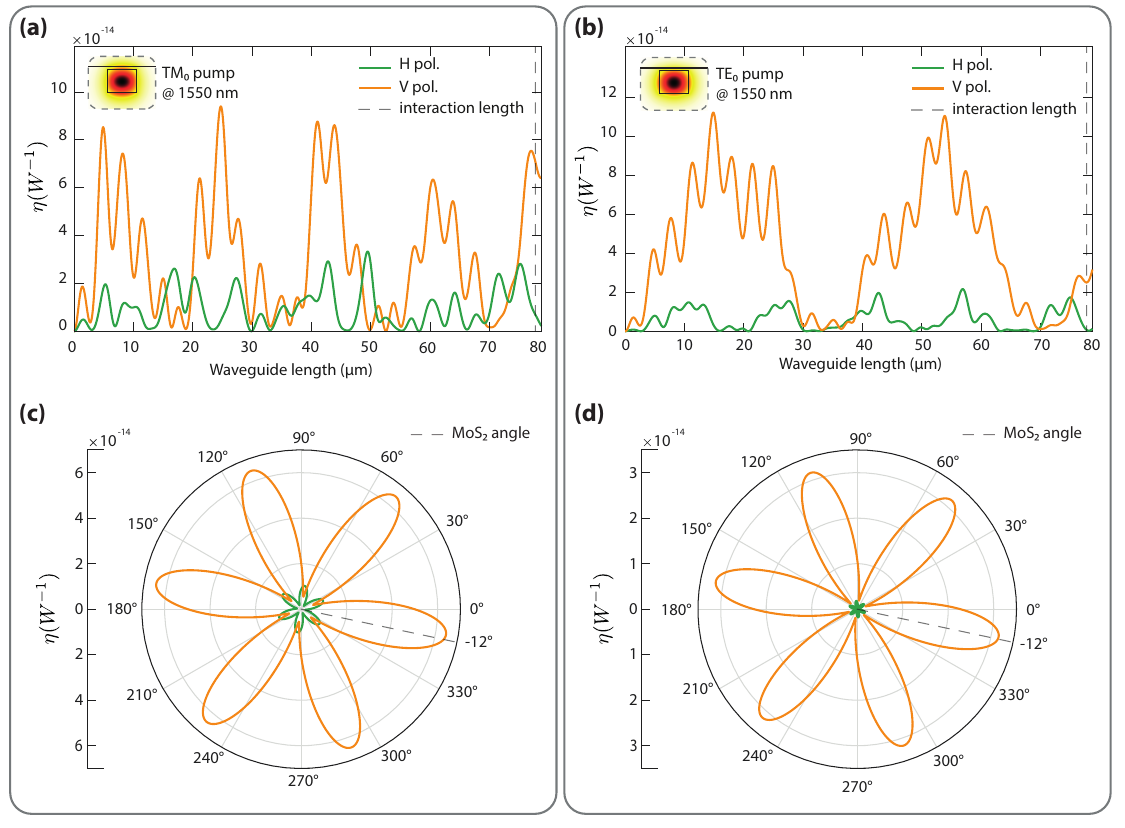}
  \caption{\textbf{Collective contribution of signal modes for the conversion efficiency:} (a-b) Overall SH-conversion efficiency as a function of the interaction length in the MoS$_2$-loaded WG for (a) TM$_{0}$ and (b) TE$_{0}$ pumping at 1550 nm, taking into account $\theta = 12^{\circ}$. The insets show the field profile of the pump mode in each case. (c-d) Overall SH-conversion efficiency as a function of the angle between the WG axis and MoS$_2$ armchair for (c) TM$_{0}$ and (d) TE$_{0}$ pumping at 1550 nm, taking into account $L$ = 79 $\mu$m. The simulations took into account the 16 signal modes. The green (orange) curve displays the conversion efficiency for the H (V) signal polarization with respect to the chip. The dashed gray lines indicate the interaction length of the MoS$_2$-loaded WG ($L$ = 79 $\mu$m) and the angle between the WG axis and MoS$_2$ armchair ($\theta = 12^{\circ}$).}
  \label{fig:collective_contribution}
\end{figure}

The SHG signal coming out of the 1 $\mu$m-width WG has contributions from all guided modes. The overall conversion efficiency captures the collective contribution of the guided signal modes as a function of the interaction length ($L$), shown in \cref{fig:collective_contribution}(a-b) for $\theta = 12^{\circ}$. 
In agreement with our experimental results, for both TM$_0$ and TE$_0$ pumping, the output SH power at V-polarization is stronger due to the dominant contribution of TM signal modes. Since phase matching is not addressed for this particular waveguide, each signal mode exhibits a different phase-mismatch factor with respect to the pump mode, so the overall conversion efficiency is a superposition of oscillatory SH-power contributions with different spatial periods, as described by \cref{eq:conversion_eff_main}.
\cref{fig:collective_contribution}(c-d) shows the dependence of the overall conversion efficiency with the MoS$_2$ armchair-to-WG angle ($\theta$) for $L = 79$ $\mu$m. At any given interaction length, the vectorial nature of the spatial overlap between the pump and dominant signal modes leads to a combination of $cos(3\theta)$ and $sin(3\theta)$ contributions to the nonlinear coefficient, which makes the angle dependence of the conversion efficiency to deviate from the expected behavior, with only either a $cos(3\theta)$ or a $sin(3\theta)$ contribution to the nonlinear coefficient (See \cref{eq:angleDependence_main}). For example, at $L = 79$ $\mu$m, the maximum conversion efficiency for both pump polarizations is shifted away from the expected angles ($\theta = m\cdot \,30^{\circ}$, $m \in \mathbb{Z}$) by approximately 10 degrees.    



\begin{figure}
  \includegraphics[width=\linewidth]{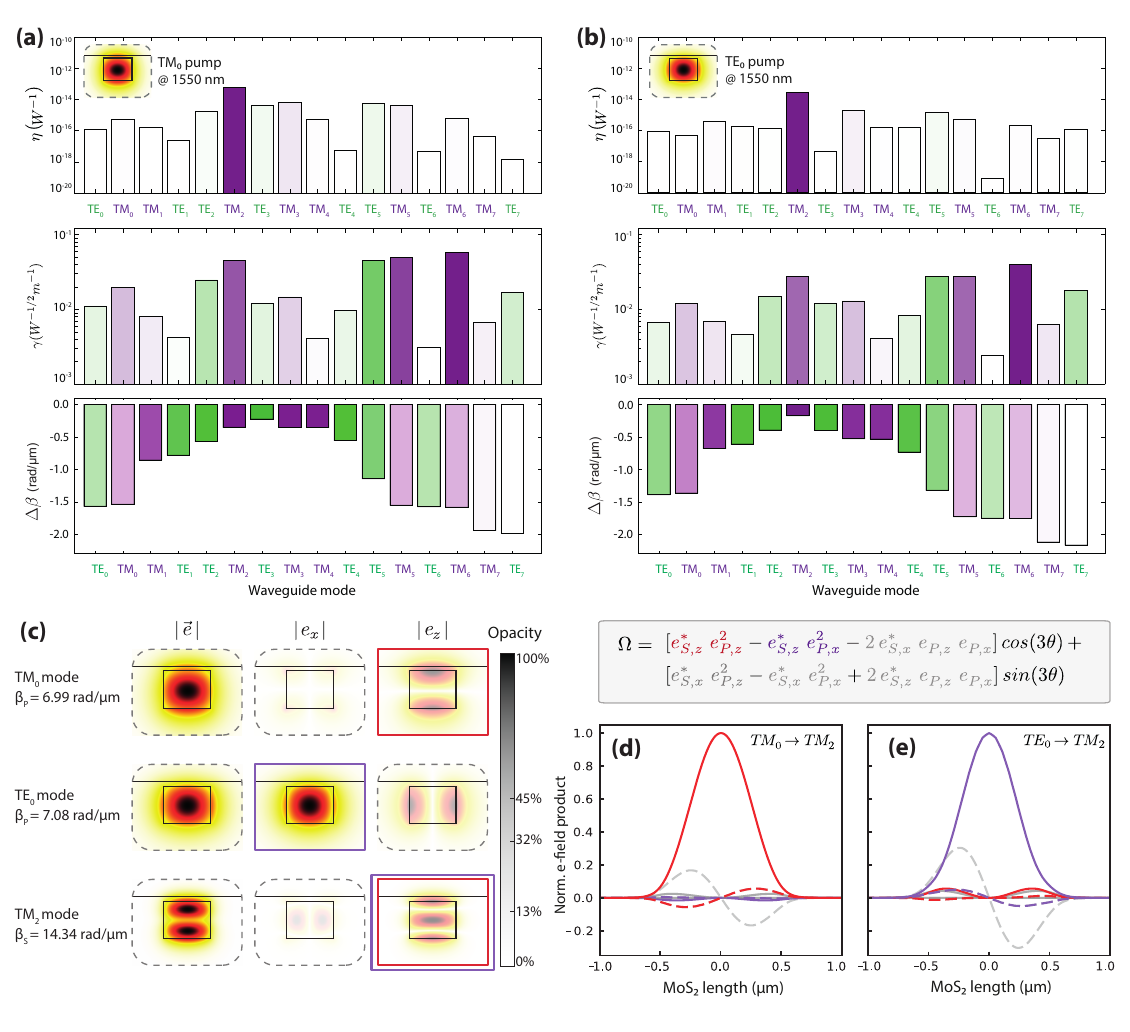}
  \caption{\textbf{Simulation results of SHG in the MoS$_2$-loaded WGs for $\theta = 12^{\circ}$ and L = 79 $\mu$m:} (a,b) Conversion efficiency ($\eta$), nonlinear coefficient ($\gamma$) and phase mismatch factor ($\Delta \beta$) for the 16 signal modes at 775 nm with (a) TM$_{0}$ and (b) TE$_{0}$ as the pump field. In both cases, the insets show the field profiles of the pump mode. The green (purple) color denotes TE (TM) modes. The color grading follows the $\eta$ and $\gamma$ scale (the higher the value, the darker the color), and the $\Delta \beta$ scale (the lower the $\Delta \beta$, the darker the color).  (c) Normalized mode profile ($|\vec{e}|$) and $x$- and $z$-components of the $e$-field for the pump modes at 1550 nm and the signal mode with the highest $\eta$ (TM$_2$). The opacity attributed to the field profiles is defined by the ratio between the maxima of the relevant electric field component ($|\vec{e_x}|$ or $|\vec{e_z}|$) and that of the mode profile ($|\vec{e}|$). (d,e) Real (solid) and imaginary (dashed) parts of the electric field products between the pump and signal modes that contribute to the overlap integral along the MoS$_2$ domain for the (d) TM$_{0}$ $\rightarrow$ TM$_{2}$ and the (e) TE$_{0}$ $\rightarrow$ TM$_{2}$ SHG interactions. The color coded \cref{eq:angleDependence_main} is indicated in the legend above the plots for reference of the field products. The electric field components that yield the highest field products - indicated in red and purple, respectively, in (d) and (e) are indicated by squares of the same colors in (c).}
  \label{fig:SHG_physics}
\end{figure}

To reveal the individual modal 
contributions to SHG in the MoS$_2$-loaded WGs, in \cref{fig:SHG_physics}(a-b) we show the phase mismatch, nonlinear coefficient and conversion efficiency for the 16 signal modes supported by the MoS$_2$-loaded WG ($L = 79$ $\mu$m, $\theta = 12^{\circ}$), for TM$_0$ and TE$_0$ pumping at 1550 nm. The signal modes with the highest conversion efficiency are those with an optimal balance between nonlinear coefficient and phase mismatch. Although several signal modes exhibit high $\gamma$, especially higher-order modes, only those with a smaller $\Delta \beta$ contribute significantly to SHG (e.g., TM$_2$, TM$_3$, and TE$_5$). Regardless of whether the pump is in the TM$_0$ or TE$_0$ mode, the signal mode with the highest conversion efficiency is a TM mode: TM$_2$. This result is surprising since the MoS$_2$ $\chi^{(2)}$ does not couple the pump fields along the out-of-plane direction. Although the TM$_{2}$ mode exhibits a negligible $x$-field component ($e_x$), its field component along the WG axis ($e_z$) plays a determinant role in SHG, since it overlaps significantly with both the pump TM$_0$ and TE$_{0}$ modes in the MoS$_2$ domain, as shown in \cref{fig:SHG_physics}(c-e). 
The individual pump-signal field product terms over the MoS$_2$ domain are shown in \cref{fig:SHG_physics}(d) and (e) for the TM$_{0}$ $\rightarrow$ TM$_{2}$ and TE$_{0}$ $\rightarrow$ TM$_{2}$ modal interactions, respectively. Note that the high conversion efficiency of the TM$_{0}$ $\rightarrow$ TM$_{2}$ (TE$_{0}$ $\rightarrow$ TM$_{2}$) interaction is driven primarily by the dominant term $e_{S,z}^*~e_{P,z}^2$ ($e_{S,z}^*~e_{P,x}^2$) of \cref{eq:angleDependence_main}. These field products are proportional to $cos(3\theta)$, resulting in maximum $\eta$ at the angles $\theta = m\; 60^{\circ}$, with $m \in \mathbb{Z}$. Cross-polarized modal interactions, such as TE $\rightarrow$ TM and TM $\rightarrow$ TE, have their conversion efficiency driven, respectively, by the dominant field product term $e_{S,z}^*~e_{P,x}^2$ and $e_{S,x}^*~e_{P,z}^2$, whereas co-polarized modal interactions, such as TE $\rightarrow$ TE and TM $\rightarrow$ TM, have their $\eta$ determined, respectively, by the field product terms $e_{S,x}^*~e_{P,x}^2$ and $e_{S,z}^*~e_{P,z}^2$. This holds true unless the dominant field product is antisymmetric (e.g., imaginary part of the field product $e_{S,z}^*~e_{P,z}^{2}$ in \cref{fig:SHG_physics}(d)), in  which case the contribution to the overlap integral becomes zero. 

\begin{figure}
  \centering
    \includegraphics[width=1\linewidth]{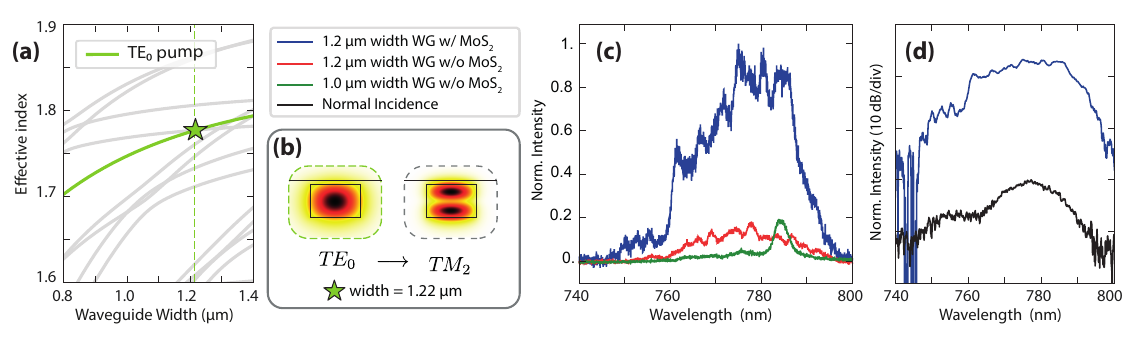}
  \caption{\textbf{Phase matching analysis in the MoS$_2$-loaded WGs:} (a) Effective index as a function of WG width for the pump TE$_0$ (green) mode at 1550 nm and signal modes at 775 nm (gray). The green star at 1.22 $\mu$m corresponds to phase matching for the TE$_{0}$ $\rightarrow$ TM$_{2}$ interaction. (b) Mode profiles (norm of the electric field) for the phase-matched modes at the width of 1.22 $\mu$m. (c) SH intensity normalized by the pump output power squared divided by the effective mode area for horizontal pump polarization and vertical SH polarization, in the 1.2\,$\mu$m-width WG with MoS$_2$ (blue), a 1.2\,$\mu$m-width WG without MoS$_2$ (red), and a 1.0\,$\mu$m-wide WG without MoS$_2$ (green). (d) Comparison between SH intensity, normalized by the pump output power squared divided by the effective mode area, as measured for normal incidence excitation of an MoS$_2$ flake (black) and for the 1.2\,$\mu$m-wide WG (blue), with the pump at 1560 nm. }
  \label{fig:dispersion_curves}
\end{figure}

The MoS$_2$-loaded WG design can be improved by tailoring of phase matching and crystal orientation, and by extending the MoS$_2$-WG interaction length. Indeed, it has been shown that large-area TMD layers can be obtained by either chemical vapor deposition (CVD) ~\cite{TMD_CVD-large} or gold-assisted exfoliation~\cite{AuExf2018,AuExf2016,Murilo2022}, which can provide $>$1 mm long single crystals. Phase matching management can be achieved via proper design of the WG's cross section. \cref{fig:dispersion_curves}(a) shows the dispersion curves for the TE$_0$ pump mode at 1550 nm and signal modes as a function of the WG width. Phase matching involving lower-order modes is achieved for a width of 1.22 $\mu$m for the interaction TE$_0$ $\rightarrow$ TM$_2$ (\cref{fig:dispersion_curves}(b)). In this case, the dominant term of the field overlap integral is $e_{S,z}^*~e_{P,x}^2$, which is proportional to $cos(3\theta)$, thus pushing the maximum conversion efficiency to $\theta = 0^{\circ}$. 

We have, thus, investigated the impact of phase-matching by measuring SHG in 1.2 µm-width, 0.8 µm-height WGs, fabricated at Ligentec, onto which we transferred an approximately 110 µm-long MoS$_2$ monolayer with the armchair direction forming an angle of 1.7 degrees with the WG axis. For comparison, we measured the SH-power for the H pump/V signal polarization configuration, for waveguides with widths of 1 $\mu$m and 1.2 $\mu$m, with and without MoS$_2$, while pumping them with a 40 fs-laser at 1560 nm, delivering 5 kW of peak power at 99 MHz of repetition rate. The total length of these waveguides was 2.5 mm. 
The corresponding results are presented in \cref{fig:dispersion_curves}(c). We observe enhancement of SHG in the phase-matched WGs with MoS$_2$ by factors $7$ and $14$, as compared to the 1.2 $\mu$m-width WG without MoS$_2$ and the 1 $\mu$m-width WG without MoS$_2$, respectively. Furthermore, the spectral profile for the 1.2 $\mu$m-width WGs exhibits small oscillations, indicating phase-matching effects. This data provides direct experimental validation of the tailored phase-matching conditions achieved through our design. The quadratic dependence of the SH-power on the pump power for the 1.2 $\mu$m-width WG can be found in the Supplementary Information section S4. Moreover, polarization-resolved SHG experiments, shown in Fig. S1(a-b), confirm that strain plays negligible role in the presented SHG results. 

We also compared SHG in the MoS$_2$-loaded phase-matched WG with that obtained with the free-space excitation of an MoS$_2$ monolayer on a flat substrate (glass slide) at normal incidence and in transmission. For a fair comparison, and in line with conventional SHG efficiency analysis (see, e.g., \cite{yariv1989quantum}), we normalised the SH intensities by their respective $P_p^2/A_p$ factor, where $P_p$ is the pump output power and $A_p$ is the pump area (either the focused pump area or the modal area, in the cases of free space and WG, respectively). The results, presented in Fig. \ref{fig:dispersion_curves}d, show that even with only 110 µm ($\sim 4\%$) of the phase-matched waveguide length covered with MoS$_2$, an enhancement by a factor $\sim220$ is attained, unequivocally highlighting the benefit of our approach.
Further improvements in SHG are expected through design engineering strategies that shift the modes closer to the MoS$_2$ flake~\cite{inga2020alumina} or by employing thicker flakes, such as MoS$_2$ 3R~\cite{zhao2016atomically,shi20173r, xu2022towards}. 
Our findings provide clear proof of the SHG enhancement achieved through the incorporation of the MoS$_2$ monolayers onto SiN WGs. They also provide important guidelines for the design of WGs integrated with monolayer TMDs, aiming at nonlinear optical frequency conversion applications.

\section{Conclusion}
We studied SHG in on-chip SiN WGs integrated with monolayer MoS$_2$. Our experimental results showed enhancement of SHG in the MoS$_2$-loaded WGs as compared to the ones without the monolayer for co- (TE $\rightarrow$ TE and TM $\rightarrow$ TM) and cross-polarized (TE $\rightarrow$ TM and TM $\rightarrow$ TE) modal interactions. In agreement with the experimental results, our simulations revealed a stronger contribution of TM signal modes (and, in particular, mode TM$_2$) to SHG, which, for the employed waveguide design, overlap the most with both pump modes TE$_0$ and TM$_0$ in the MoS$_2$ domain. Our analysis took into account the full vectorial nature of the modal fields and the tensorial nature of the MoS$_2$’s second-order susceptibility.
Although monolayer MoS$_2$'s $\chi^{(2)}$ does not couple pump fields to out-of-plane signal fields, the electric field component of the TM$_2$ signal mode along the WG axis ($e_z$) strongly overlaps with the pump fields over the MoS$_2$ domain, increasing the conversion efficiency for this particular mode. These interactions cannot be predicted with a scalar model that takes into account only the main electric field component of each mode, i.e., in-plane $e_x$ for a TE mode and out-of-plane $e_y$ for a TM mode. Finally, by means of modal phase matching for the TE$_0$ $\rightarrow$ TM$_2$ interaction (obtained via simulation-assisted WG design), we obtained $220\times$ conversion enhancement relative to free space (normal incidence) excitation, as well as $14\times$ enhancement relative to the non-phase-matched case, with only 110 µm of pump-MoS$_2$ interaction length.

Beyond its direct applicability to the SiN–MoS$_2$ system studied here, the full vectorial–tensorial framework developed in this work is material-agnostic and can be readily extended to other nonlinear 2D materials and hybrid architectures. By providing the corresponding second-order susceptibility tensor and refractive-index dispersion, the same model can accurately describe SHG interactions in other TMDs (e.g., WS$_2$, WSe$_2$) as well as in non-TMD materials such as ReS$_2$ or elemental tellurium. This versatility makes the model a general predictive tool for designing and optimizing a wide range of 2D-material–integrated nonlinear photonic platforms. Our findings offer valuable insights into the nonlinear modal interaction and the design of on-chip Si-photonic devices integrated with 2D materials towards second-order nonlinear optical applications, such as nonlinear frequency conversion, generation of entangled photon pairs and parametric amplification.


\section*{Methods}
\label{sec:methods}
\threesubsection{MoS$_2$-loaded waveguide design and fabrication}
MoS$_2$ monolayers were exfoliated from bulk 2H-phase MoS$_2$ bulk crystal (2D Semiconductors) by the standard scotch tape technique onto a viscoelastic polydimethylsiloxane (PDMS) stamp for inspection under an optical microscope. The SiN WGs used in this work were designed by us and fabricated by Ligentec SA. Scanning electron microscopy images of the WGs are provided in the Supplementary Information section 3. The SiN core of the WGs is cladded in a SiO$_2$ substrate and has cross-section dimensions of 1 $\mu$m $\times$ 0.8 $\mu$m (width $\times$ height). As schematically shown in \cref{fig:SHG_experiment}a, the WG is L shaped, to prevent stray light from being collected at the output. Close to the input and output, the SiO$_2$ thickness over the WG is of 6.3 µm, being reduced to 100 nm where the MoS$_2$ flake is to be deposited, which is the standard SiO$_2$ thickness set by Ligentec for regions of the chip where the waveguide’s top surface must be exposed. There is an inverted taper at the chip input to optimize light coupling to the TE$_0$ mode in the waveguide. The inverted taper is a feature available in the foundry’s process design kit (PDK), whose specifications are not shared with customers. There is no inverted taper at the output, so that all SH output modes are efficiently collected and analyzed. The selected MoS$_2$ flake was aligned and transferred onto the SiN WG at 60°C by using a commercial transfer system (HQ Graphene). 
\cref{fig:design_considerations}(d) shows the optical image of a representative MoS$_2$ flake transferred to the WG.

\threesubsection{Raman characterization}
We employed high-resolution Raman spectroscopy to confirm the monolayer nature of the transferred flakes ~\cite{CadoreRaman_2024}. The measurements were conducted using a WITec 300R spectrometer with a 532 nm excitation laser operating at a power below 1 mW to prevent sample damage. A confocal micro-configuration was utilized with a 100x (NA = 0.95) microscope objective lens, an xyz stage, and an 1800 lines/mm grating at room temperature. The vibrational modes were labeled using the irreducible representations of the D$_{6h}$ point group, applicable to bulk MoS$_2$. Group-theoretical analysis predicts four Raman-active modes for the D$_{6h}$ group ~\cite{verble1970lattice}: three in-plane modes E$^1_g$, E$^1_{2g}$, and E$^2_{2g}$, and one out-of-plane mode A$^1_g$. In our experimental configuration, only E$^1_{2g}$ and A$^1_g$ are observed as E$^2_{2g}$ mode lies at very low frequencies ($\sim$ 30 cm$^{-1}$) ~\cite{XuMoS2Raman}, while the E$^1_g$ mode is forbidden in backscattering geometry on the basal plane ~\cite{frey1999raman}. The separation between the E$^1_{2g}$ and A$^1_g$ modes decreases with reducing material thickness, serving as a reliable indicator for the number of MoS$_2$ layers ~\cite{Tywoniuk2025}. 

\threesubsection{Polarization-resolved SHG}
To confirm the flakes’ alignment, we performed polarization-resolved SHG ~\cite{CadoreJAP2024}. For this, a linearly polarized  pulsed laser at 1560 nm (mode-locked Erbium-doped fiber laser delivering 150-fs pulses at a repetition rate of 89 MHz) was focused on the MoS$_2$ flake on top of the WG at normal incidence, and the reflected light was sent through an analyzer and directed to a spectrometer (Andor Kymera coupled to a silicon CCD iDus 416 camera). The maximum (minimum) SHG intensity occurs when the laser polarization is parallel to the armchair (zigzag) crystal axis~\cite{MalardPRB213, woodward2016characterization}. By collecting spectra at the SH-wavelength while simultaneously rotating the laser polarization and the analyzer axis in steps of 5 degrees, we obtained the polar plots shown in Fig. S1(a-b), that reveals -12 degree and 1.7 degree angles between the MoS$_2$ armchair direction and the WG axis, for the 1 µm-width and 1.2 µm-width WGs, respectively.

\threesubsection{Numerical simulations}
The field overlap integrals of \cref{eq:gamma_main} for a SiN WG integrated with a MoS$_2$ monolayer, as shown in \cref{fig:design_considerations}(c), were calculated in Comsol Multiphysics. The field components of the pump and signal modes at 1550 and 775 nm, respectively, were calculated through a mode analysis study. Once the field solutions were found, they were projected onto a shared 2D geometry where the WG cross-section is defined, so that their field overlap integrals could be performed. 
The dispersion contribution and propagation losses induced by the MoS$_2$ monolayer were neglected in the simulation.
A script was implemented to calculate the nonlinear coefficient and the conversion efficiency considering the pump mode as TE$_{0}$ and TM$_{0}$ modes and the first 16th signal modes calculated in Comsol. The second-order susceptibility considered in the calculations was $2.0\times10^{-20}$ m$^2$/V \cite{woodward2016characterization}.

\threesubsection{Nonlinear optical characterization of the MoS$_2$-loaded waveguides}
The experimental characterization of SHG was carried out by pumping the 1 $\mu$m-width MoS$_2$-loaded WGs ($10$~dB of insertion loss) with an input (off-chip) peak power of 750 W generated by a mode-locked Erbium-doped fiber laser at 1560 nm, delivering 150-fs pulses at a repetition rate of 89 MHz. The linear polarization of the pump was adjusted with the aid of a half-wave plate and a polarizer to be either horizontal or vertical, enabling the excitation of TE or TM modes in the WG, respectively. A microscope objective lens (Newport MV-40X; NA = 0.65) was utilized to couple the pump light into the WGs. Light at the WG output was collected by a 10x objective lens (Mitutoyo M Plan Apo-10x; NA = 0.28)
and directed to either a refrigerated spectrometer (Solis, Andor) or a power meter for analysis. The acquisition time for the SHG collection was set to 5 seconds. To obtain the results of \cref{fig:dispersion_curves}(c-d), 1.2 $\mu$m-width (with and withoud MoS$_2$) and 1 $\mu$m-width (without MoS$_2$) WGs ($12$~dB of insertion loss) were pumped by an input (off-chip) peak power of 5 kW generated by another mode-locked Erbium-doped fiber laser at 1560 nm, delivering 40-fs pulses at a repetition rate of 99 MHz.

\medskip
\textbf{Supporting Information} \par 
The Supplementary Material provides the theoretical description of second harmonic generation in the MoS$_2$-loaded waveguides, experimental characterization of the MoS$_2$ flakes and of the waveguides, as well as complementary on-chip SHG characterization data.

\medskip
\textbf{Data Availability}

Data underlying the results presented in this paper are not publicly available at this time but may be obtained from the authors upon reasonable request.

\medskip
\textbf{Disclosures}

The authors declare no conflicts of interest.

\medskip
\textbf{Acknowledgements} \par 
This work is supported by the São Paulo Research Foundation – FAPESP (grant nos. 2020/04686-8, 2022/07892-3, 2020/04374-6, 2018/15577-5 and 2018/25339-4), CNPq (grant nos. 309920/2021-3, and 141698/2023-3), Provost Office of Research and Innovation of University of Sao Paulo (grant no. 2022.1.9345.1.2), the Brazilian Nanocarbon Institute of Science and Technology (INCT/Nanocarbono), and MackPesquisa. All authors are also thankful to the Brazilian Nanotechnology National Laboratory (LNNano) and Brazilian Synchrotron Light Laboratory (LNLS), part of the Brazilian Centre for Research in Energy and Materials (CNPEM), a private non-profit organization under the supervision of the Brazilian Ministry for Science, Technology, and Innovations (MCTI), for sample preparation – Proposals: MNF-20240167 and LAM-2D-20232114.

\medskip


\bibliographystyle{MSP}
\bibliography{MoS2paper_ACSPhoton.bib}

\newpage
\renewcommand{\theequation}{S\arabic{equation}}
\renewcommand{\thesection}{S\arabic{section}}
\renewcommand{\thesubsection}{\Alph{subsection}}
\renewcommand{\thesubsubsection}{\roman{subsubsection}}
\renewcommand{\thefigure}{S\arabic{figure}}
\renewcommand{\thetable}{S\arabic{table}}
\setcounter{figure}{0}
\setcounter{table}{0}
\setcounter{equation}{0}
\setcounter{section}{0}

\end{document}


\section*{Supplementary Information for: Vectorial engineering of second-harmonic generation in silicon-based waveguides integrated with 2D materials}

\author{Mohd Rehan$^{\S}$}
\author{Nathalia B. Tomazio$^{\S}$}
\author{Alisson R. Cadore}
\author{Daniel F. Londono-Giraldo}
\author{Daniel A. Matos}
\author{Gustavo S. Wiederhecker}
\author{Christiano J. S. de Matos$^*$}

\begin{affiliations}
$\S$ Authors who contributed equally to this work.

M. Rehan, D. F. Londono-Giraldo, C. J. S. de Matos\\
School of Engineering, Mackenzie Presbyterian University, Sao Paulo 01302-907, SP, Brazil\\
MackGraphe, Mackenzie Presbyterian Institute, Sao Paulo 01302-907, SP, Brazil\\
Email Address: cjsdematos@mackenzie.br

N. B. Tomazio, D. A. Matos\\
Instituto de Física, Universidade de São Paulo, São Paulo, Brazil

A. R. Cadore\\
Brazilian Nanotechnology National Laboratory (LNNano), Brazilian Center for Research in Energy and Materials (CNPEM), Campinas 13083-100, SP, Brazil\\
Programa de Pós-Graduação em Física, Instituto de Física, Universidade Federal de Mato Grosso,
Cuiabá 78060-900, MT, Brazil

G. S. Wiederhecker\\
Photonics Research Center, Gleb Wataghin Physics Institute, Universidade Estadual de Campinas (UNICAMP), Campinas, SP, Brazil
\end{affiliations}

\section{Theoretical aspects of SHG in MoS$_2$-loaded waveguides}
\label{sec:theo_aspects_SHG}

To reveal the modes that contribute the most with SHG in the silicon nitride (SiN) waveguides (WGs) integrated with MoS$_2$ monolayer, and determine the proper orientation of the MoS$_2$ crystallographic axes with respect to the WG, we modelled the SHG process with nonlinear coupled equations in the slowly varying amplitude approximation (SVEA) \cite{boyd2020nonlinear,shen1984principles, chen2019tunable}. In this treatment, the fields, at a given frequency $\omega$, are described in the waveguide following a mode expansion strategy:

\begin{equation}
    \vec{E} = \sum_{j} a_{j}(z) \, \vec{e}_{j}(x,y)\,\exp{(i \beta_j z)},
	\label{eq:e-field_nonlinear0}
\end{equation}

\noindent where $a_{j}(z)$, $\vec{e}_{j}(x,y)$ and $\beta_j$ are, respectively, the slowly varying envelope, the field profile and the propagation constant of the $j$-th mode. The index $j$ refers to either pump or signal modes. The fields are described in the WG reference frame, so the coordinates $z$ and $x$/$y$, respectively, represent the WG direction and its cross-sectional directions. By taking the mode expansion description of the fields into Maxwell's equations, we can derive the equation that governs the evolution of the $j$-amplitude along the WG axis:

\begin{equation}
	\frac{\partial a_{j}}{\partial z} = i\frac{\,\omega_{j}}{4} \left[\int \vec{e}\,^*_{j} \cdot \vec{P}_{j}^{}\,da \right] \exp{(-i\,\beta_{j} z)},
	\label{eq:e-field_nonlinear1}
\end{equation}



\noindent where $\vec{P}_{j}$, $\omega_{j}$ and $\beta_{j}$ are, respectively, the nonlinear polarization, angular frequency and propagation constant of the $j$-mode. The summation symbols were omitted for clarity. 
\noindent The mode field profiles ($\vec{e}_{j}$, $\vec{h}_{j}$) are normalized so that the power transmitted by the $j$-mode is equal to the amplitude squared ($|a_{j}|^2$): 

\begin{equation}
    p_{j} =|a_{j}|^2\, \frac{1}{2}\,\int \vec{e_{j}}^* \times \vec{h}_{j}\cdot \hat{z}\, da \equiv |a_{j}|^2,
\end{equation}

\noindent which implies: 
 
 \begin{equation}
    \vec{e}_{j} =\frac{\tilde{\vec{e}}_{j}}{\sqrt{\frac{1}{2} \int \tilde{\vec{e_{j}}^*} \times \tilde{\vec{h}}_{j}\cdot \hat{z}\, da}},
\end{equation}

 \begin{equation}
    \vec{h}_{j} =\frac{\tilde{\vec{h}}_{j}}{\sqrt{\frac{1}{2} \int \tilde{\vec{e_{j}}^*} \times \tilde{\vec{h}}_{j}\cdot \hat{z}\, da}},
\end{equation}

\noindent where $\tilde{\vec{e}}_{j}$ ($\tilde{\vec{h}}_{j}$) is the electric (magnetic) field of the $j$-th mode in V/m (A/m) units. The index $j$ refers to either the pump or signal field.  


The components of the second-order polarization are described by \cite{boyd2020nonlinear}: 

\begin{equation}
    P_l(2\omega) = \varepsilon_0 \sum_{m,n} \chi_{lmn}^{(2)}(2\omega; \omega, \omega) E_m(\omega) E_n(\omega),
	\label{eq:nonlinear polarization}
\end{equation}

\noindent in which the indices $l$, $m$ and $n$ refer to the cartesian components of the fields, $\varepsilon_0$ denotes the vacuum permittivity, and $\chi_{lmn}^{(2)}$ denotes the second-order susceptibility tensor. More specifically, $\chi_{lmn}^{(2)}$ is the susceptibility component that couples the input fields with $m$ and $n$ polarizations into the nonlinear polarization $l$. 
Following convention, we have written $\chi_{lmn}^{(2)}$ as a function of three frequency arguments, where the first argument is the sum of the other two.

Given that the crystalline structure of MoS$_2$ monolayer belongs to the symmetry group $\mathcal{D}_{3h}$~\cite{malard2013observation}, it can be shown that its second-order susceptibility tensor yields only four nonvanishing and interdependent components \cite{weismann2016theoretical}:

\begin{equation}
    \chi^{(2)} := \chi^{(2)}_{z'z'z'} = -\chi^{(2)}_{z'x'x'} = -\chi^{(2)}_{x'z'x'} = -\chi^{(2)}_{x'x'z'},
	\label{eq:chi2}
\end{equation}

\noindent where $z'$ and $x'$ are, respectively, the armchair and the orthogonal zigzag directions of the MoS$_2$ monolayer, as indicated in Fig. 1(c). Taking the symmetry properties of MoS$_2$ monolayer into account, the second-order polarization becomes: 

\begin{equation}
    \vec{P}=
    \begin{bmatrix}
   P_{z'} \\
   P_{x'} \\
   P_{y'} 
  \end{bmatrix} = 
  \begin{bmatrix}
   \varepsilon_0 \;\chi^{(2)}\;(E_{P,z'}^2-E_{P,x'}^2) \\
   -2\;\varepsilon_0 \;\chi^{(2)}\;E_{P,z'}\,E_{P,x'} \\
   0
  \end{bmatrix},
    \label{eq:SO-Pol}
\end{equation}

\noindent where $E_{P,z'}$($E_{P,x'}$) is the pump E-field in the $z'$($x'$) direction.
\noindent Note that there is no susceptibility component that couples the fields to the out-of-plane direction ($y'$), so the nonlinear polarization component in this direction is zero. In general, the crystal axes, $z'$ and $x'$, are not aligned to the WG axes, $z$ and $x$. The field profiles in the WG reference frame can be obtained by performing a rotation operation:  

\begin{equation}
    \begin{bmatrix}
   E_{P,z} \\
   E_{P,x}
  \end{bmatrix} = 
  \begin{bmatrix}
   cos\theta & -sin\theta \\
   sin\theta & cos\theta
  \end{bmatrix}
   \begin{bmatrix}
   E_{P,z'} \\
   E_{P,x'}
   \end{bmatrix},
\end{equation}

\noindent with $\theta$ being the angle between both reference frames, as indicated in Fig. 1(c). In the WG reference frame, the equation that describes the evolution of the amplitude of an signalmode ($j = S$ in \cref{eq:e-field_nonlinear1}) becomes: 

\begin{equation}
	\frac{\partial a_{S}}{\partial z} = i\gamma\,a_P^2 \,exp{(i\,\Delta\beta\,z)},
	\label{eq:e-field_nonlinear10}
\end{equation}

\noindent where $\gamma$ represents the nonlinear coefficient, which essentially depends on the second-order susceptibility and spatial overlap between pump and signal modes:

\begin{equation}
	\gamma = \frac{\,\omega_{S}\,\varepsilon_0\,\chi^{(2)}}{4}\,\left[\int_{MoS_2}\Omega\,da \right]\,,
	\label{eq:gamma}
\end{equation}


\noindent The terms for the field product are condensed in the variable $\Omega$, which encompass the angle dependence of the SHG interaction:

\begin{equation}
\begin{split}
\Omega = \,&[e_{S,z}^*~e_{P,z}^2 - e_{S,z}^*~e_{P,x}^2 - 2\,e_{S,x}^*~e_{P,z}^{}~e_{P,x}^{}]\,cos(3\theta)\,+\\
            &[e_{S,x}^*~e_{P,z}^2 - e_{S,x}^*~e_{P,x}^2 + 2\,e_{S,z}^*~e_{P,z}^{}~e_{P,x}^{}]\,sin(3\theta),
\end{split}
    \label{eq:angleDependence}
\end{equation}

\noindent $\Delta\beta = 2\,\beta_P-\beta_S$ represents the phase mismatch factor between the pump and signal modes. Note that the integral in \cref{eq:gamma} is taken along the MoS$_2$ monolayer domain.  

The conversion efficiency in the undepleted pump regime is obtained by integrating \cref{eq:e-field_nonlinear10} over an interaction length $L$ and taking the ratio between the pump and signal-squared powers: 

\begin{equation}
    \eta = \frac{|a_{S}|^2}{|a_P|^4} = \gamma^2\,L^2\,\frac{sin^2(\Delta\beta\,L/2)}{(\Delta\beta\,L/2)^2}
    \label{eq:conversion_eff}
\end{equation}

\section{Characterization of the MoS$_2$ flakes}
\begin{figure}[h]
\begin{center}
  \includegraphics[width=0.9\linewidth]{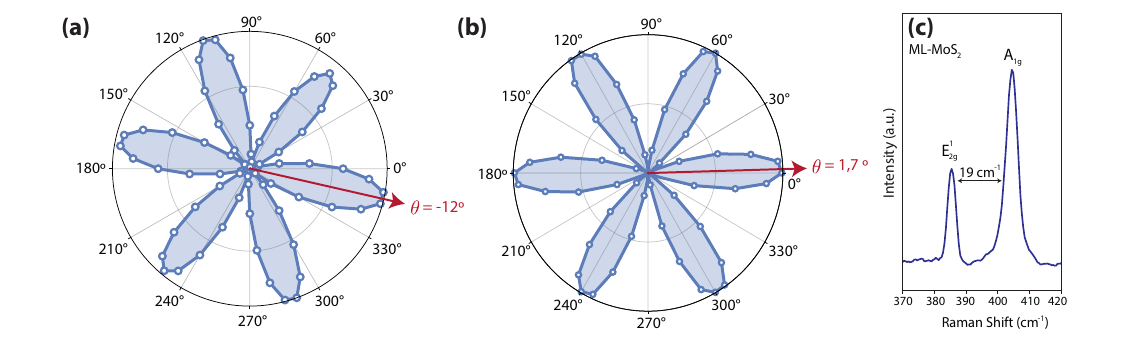}
  \caption{\textbf{Characterization of the MoS$_2$ flakes:} (a-b) Polar plots of the polarization dependent normalized SH intensity (co-polarized pump:SH configuration) for the MoS$_2$ monolayers transferred onto the (a) 1 $\mu$m-width WG and the (b) 1.2 $\mu$m-width WG, obtained under normal incidence excitation. (b) Raman spectrum collected at the MoS$_2$ flake on top of the 1 $\mu$m-width WG under 532 nm laser excitation.} 
  \label{fig:characterization_MoS2}
\end{center}
\end{figure}

To confirm the orientation of the MoS$_2$ flake with respect to the waveguide axis, we carried out polarization-resolved SHG experiments, where the reflected second harmonic intensity is collected while the pump laser polarization is rotated, upon normal incidence excitation of the MoS$_2$ flake. The same broadband polarizer is used to polarize the incident pump and the reflected SH, which thus ensures that these waves are co-polarized. \textit{See Methods} for more details. The polarization-resolved SHG measurements for the 1 $\mu$m- and 1.2 $\mu$m-width WGs are shown in \cref{fig:characterization_MoS2}a and b, respectively, revealing an angle of - 12 and 1.7 degrees between the WG axis and MoS$_2$ armchair direction, respectively.  

The SH intensity dependence on the pump polarization is affected by the presence of uniaxial strain in MoS$_2$ \cite{mennel2019second}. In the absence of uniaxial strain, the dependence on the pump polarization angle corresponds to a sine curve with a 60$^\circ$ period, which in a polar plot corresponds to a 6-petalled pattern, with petals exhibiting the same amplitude. The polar plots for the 1-$\mu$m- and 1.2-$\mu$m-width WGs (\cref{fig:characterization_MoS2}(a-b)) demonstrate only a small asymmetry in the amplitude of the petals, which indicates negligible uniaxial strain. We can, thus, conclude that strain plays a minimal role in the presented results.

The Raman spectrum measured at the MoS$_2$ flake on top of the 1 $\mu$m-width WG is shown in \cref{fig:characterization_MoS2}(c), confirming its monolayer nature.   

\newpage\section{Characterization of the waveguides}

\begin{figure}
\begin{center}
  \includegraphics[width=0.65\linewidth]{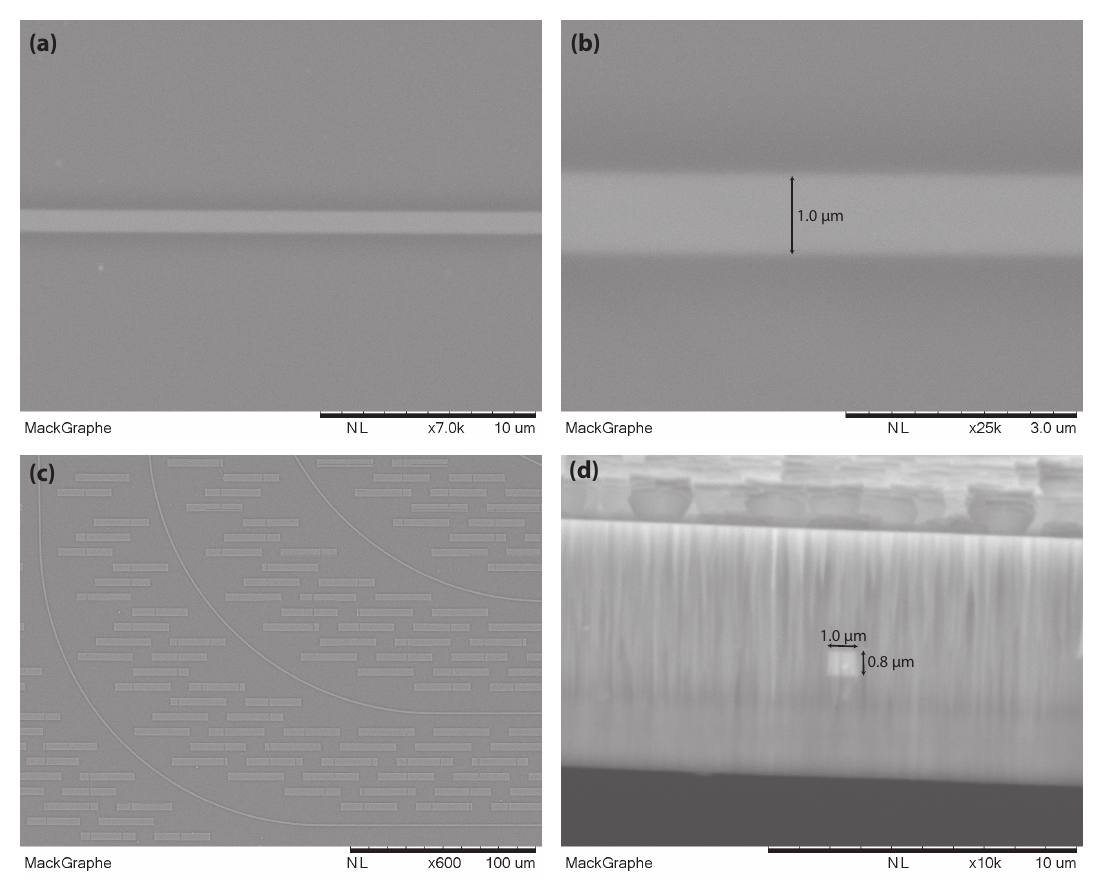}
  \caption{\textbf{SEM images of the on-chip WGs:} (a) Zoom-out and (b) zoom-in top views of the chip region where the top surface of the WGs is exposed; (c) curved part of the L-shaped WGs; (d) cross-sectional view of the chip output facet.} 
  \label{fig:mev}
  \end{center}
\end{figure}

For a more detailed characterization of the waveguide geometry, we have performed scanning electron microscope (SEM) imaging of the fabricated chips (\cref{fig:mev}). The SEM image in \cref{fig:mev}(d) corresponds specifically to the output facet of the waveguide, which, as designed, does not include the inverted taper. An inverted taper exists only in the input side of the waveguide, where it improves coupling efficiency  to the pump TE$_0$ mode. However, this taper cannot be imaged from the top, because the entire input region is covered by a 6.3-$\mu$m-thick SiO$_2$ upper cladding layer. In contrast, the region where the 2D material was transferred has a thin SiO$_2$ layer (100 nm) on top of the waveguide, allowing for efficient overlap between the optical mode and the MoS$_2$ monolayer. 
The SEM images confirm that the exposed waveguide core has a 1-$\mu$m width and 0.8-$\mu$m height. We additionally note that the devices were fabricated by Ligentec (Switzerland), ensuring reproducibility and uniformity of both the structural and optical properties across the chip.

\newpage\section{Quadratic dependence of SH power on pump power}

As shown in Fig. S3, the output power coming out of the waveguides at 780 nm shows a quadratic dependence on the pump power, as expected for SHG.

\begin{figure}[h]
\begin{center}
  \includegraphics[width=1\linewidth]{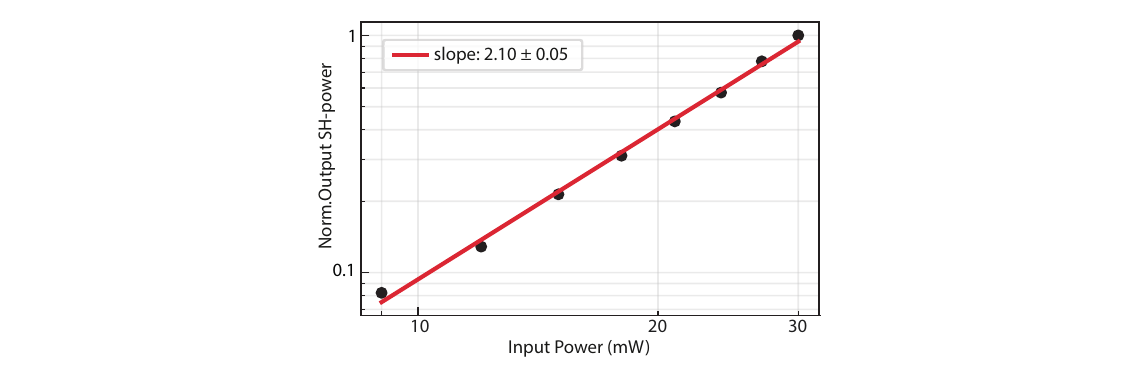}
  \caption{\textbf{Quadratic dependence of SH power on pump power:} Integrated output SH intensity at 780 nm as a function of pump power for the 1.2 $\mu$m-width waveguide, pumped with 40 fs-laser delivering 5 kW of peak power at 1560 nm and 99 MHz of repetition rate. For this measurement, the input and output polarizations were set to horizontal and vertical, respectively.} 
  \label{fig:linear_dependence}
\end{center}
\end{figure}

\newpage
\bibliographystyle{MSP}
\bibliography{MoS2paper_ACSPhoton.bib}